\PassOptionsToPackage{unicode}{hyperref}
\PassOptionsToPackage{hyphens}{url}
\documentclass[
]{article}
\usepackage{lmodern}
\usepackage{amsmath}
\usepackage{ifxetex,ifluatex}
\ifnum 0\ifxetex 1\fi\ifluatex 1\fi=0 
  \usepackage[T1]{fontenc}
  \usepackage[utf8]{inputenc}
  \usepackage{textcomp} 
  \usepackage{amssymb}
\else 
  \usepackage{unicode-math}
  \defaultfontfeatures{Scale=MatchLowercase}
  \defaultfontfeatures[\rmfamily]{Ligatures=TeX,Scale=1}
\fi
\IfFileExists{upquote.sty}{\usepackage{upquote}}{}
\IfFileExists{microtype.sty}{
  \usepackage[]{microtype}
  \UseMicrotypeSet[protrusion]{basicmath} 
}{}
\makeatletter
\@ifundefined{KOMAClassName}{
  \IfFileExists{parskip.sty}{%
    \usepackage{parskip}
  }{
    \setlength{\parindent}{0pt}
    \setlength{\parskip}{6pt plus 2pt minus 1pt}}
}{
  \KOMAoptions{parskip=half}}
\makeatother
\usepackage{xcolor}
\IfFileExists{xurl.sty}{\usepackage{xurl}}{} 
\IfFileExists{bookmark.sty}{\usepackage{bookmark}}{\usepackage{hyperref}}
\hypersetup{
  pdftitle={Real-Time Detection of Volatility in Liquidity Provision},
  pdfauthor={Matt Brigida},
  hidelinks,
  pdfcreator={LaTeX via pandoc}}
\usepackage[margin=1in]{geometry}
\usepackage{longtable,booktabs}
\usepackage{etoolbox}
\makeatletter
\patchcmd\longtable{\par}{\if@noskipsec\mbox{}\fi\par}{}{}
\makeatother
\IfFileExists{footnotehyper.sty}{\usepackage{footnotehyper}}{\usepackage{footnote}}
\makesavenoteenv{longtable}
\usepackage{graphicx}
\makeatletter
\def\maxwidth{\ifdim\Gin@nat@width>\linewidth\linewidth\else\Gin@nat@width\fi}
\def\maxheight{\ifdim\Gin@nat@height>\textheight\textheight\else\Gin@nat@height\fi}
\makeatother
\setkeys{Gin}{width=\maxwidth,height=\maxheight,keepaspectratio}
\makeatletter
\def\fps@figure{htbp}
\makeatother
\providecommand{\tightlist}{%
  \setlength{\itemsep}{0pt}\setlength{\parskip}{0pt}}
\ifluatex
  \usepackage{selnolig}  
\fi
\newlength{\cslhangindent}
\newlength{\csllabelwidth}
\newenvironment{CSLReferences}[3] 
 {
  \setlength{\parindent}{0pt}
  \ifodd #1 \everypar{\setlength{\hangindent}{\cslhangindent}}\ignorespaces\fi
  \ifnum #2 > 0
  \setlength{\parskip}{#2\baselineskip}
  \fi
 }%
 {}
\usepackage{calc} 

\title{Real-Time Detection of Volatility in Liquidity Provision}
\author{Matt Brigida\footnote{SUNY Polytechnic Institute, 100 Seymour
  Rd., Utica, NY 13502,
  \href{mailto:matthew.brigida@sunypoly.edu}{\nolinkurl{matthew.brigida@sunypoly.edu}}}}
\date{November 5, 2020}

\begin{document}
\maketitle
\begin{abstract}
Previous research has found that high-frequency traders will vary the
bid or offer price rapidly over periods of milliseconds. This is a
benefit to fast traders who can time thier trades with microsecond
precision, however it is a cost to the average market participant due to
increased trade execution price uncertainty. In this analysis we attempt
to construct real-time methods for determining whether the liquidity of
a security is being altered rapidly. We find a four-state Markov
switching model identifies a state where liquidity is being rapidly
varied about a mean value. This state can be used to generate a signal
to delay market participant orders until the price volatility subsides.
Over our sample, the signal would delay orders, in aggregate, over 0 to
10\% of the trading day. Each individual delay would only last tens of
milliseconds, and so would not be noticable by the average market
participant.
\end{abstract}



\clearpage

The goal of this analysis is to construct methods to determine, in
real-time, when the volatility of the liquidity provided is being
rapidly changed around a mean value, which is consistent with the effect
of an algorithm or set of algorithms. Such methods would allow the
creation of orders which can be canceled, or delayed, if the market
switches to such a regime with unstable liquidity. This is analogous to
the crumbling quote signal from the Investors Exchange (outlined in
Bishop (2017)).

Such real-time detection is a difficult task, though identification does
not have to be perfect. The threshold is that investors choose to use
the order---that it is correlated enough with undesirable activity that
it adds value to the investor to submit the order type. For the order
type to have worth to investors algorithmic activity, or other processes
which rapidly change liquidity around a mean value, which is a cost to
the average investor must exist.

Hasbrouck (2018) found evidence for substantial volatility in the bid
and offer prices which was not due to fundamental changes in the asset
value. The cost of this volatility is not borne equally by traders.
Faster traders are able to choose the point (in microseconds) at which
they trade. Slower traders, however, will receive a trade price some
time later (maybe seconds) after they attempt to submit a marketable
order. This trade price is a random variable, and they are exposed to
price risk which is a function of the expected variation of the bid (or
offer) price over the time from when they submitted the order to when it
is matched by the exchange.

So when fast traders change the bid/ask price quickly, slower traders
still \emph{expect} to receive/pay the same amount for each sell/buy
order, however they have increased uncertainty. This increased risk
without increased compensation should be avoided by any rational
investor. The goal of our analysis is to help investors find ways to
delay their order until the execution price of their order has more
certainty. Since the volatility can occur in milliseconds, the method of
identification must itself be algorithmic.

Note, investors should attempt to avoid these periods of increased
uncertainty even if the source of the uncertainty is not high-frequency
traders. We therefore don't attempt to determine the source of
uncertainty, but rather, in real time, identify when such variations in
liquidity are occurring.

Both spread and depth pose substantial risk, particularly for
institutional investors who tend to trade in qualtities far larger than
what is available at the inside quotes. Despite this many seminal models
of market making under asymmetric information ignore market depth by
assuming a unit size for all trades (Copeland and Galai (1983); Glosten
and Milgrom (1985); Easley (1992)). Alternatively, in Kyle (1985) market
depth is implicitly incorporated in the model through requiring
specialists to supply complete pricing functions. In our analysis we
will consider the time-series of liquidity available in the orderbook
within a set distance from the bid-ask midpoint.

Our algorithm will attempt to filter out the other various drivers of
price and market depth changes. For example, French and Roll (1986)
found evidence that stock price volatility is driven by private
information being incorporated into market prices via trading. Lee,
Mucklow, and Ready (1993) studied the relationship between spreads and
depth around earnings announcements. So we are attempting to find a
state where price and market depth are changing in a manner inconsistent
with trading on private information or around events. Notably, this
first source of price and depth change would impart a directional bias
to prices, and in the case of Lee, Mucklow, and Ready (1993) the spread
widened. Alternatively, the high-frequency trading we are attempting to
identify does not change \emph{mean} price or market depth as in these
former cases.

\hypertarget{data}{%
\section{Data}\label{data}}

We use data for the heavily traded E-Mini S\&P 500 Futures contract.
Price discovery in the equity market occurs in this contract (Hasbrouck
(2003)). Trading hours from Sunday--Friday from 6:00 p.m. to 5:00 p.m.
Eastern Time (ET). Contract value is \$50 times the futures price. Cash
delivery with expirations every 3 months. Traded on the Chicago
Mercantile Exchange (CME) (pit and electronic (Globex)).

The reason we use CME Data ES is because, in addition to being the first
place that information is incorporated into prices and trading
overnight, all trades and quotes take place in this one central book. So
there is no delay in orders due to location.

Data are Market Depth Data\footnote{\url{http://www.cmegroup.com/confluence/display/EPICSANDBOX/Market\%2BDepth}}
for E-Mini S\&P 500 futures (Globex), for the trading week from November
7 to November 11, 2016. The data were purchased directly from the CME.
We focus our results on November 9 2016 because it was the trading day
where results of the US Presidential election were released, and
therefore there were high levels of trade and quote volume, which makes
the presence of algorithmic activity more likely.

Market Depth Data contains all market messages (trade/limit order
updates) to and from the CME, and is time-stamped to the nanosecond. The
data also includes tags for aggressor side. Using this data we can
recreate the ES orderbook with nanosecond resolution and up to 10 levels
deep. The data are encoded in the CME's FIX/FAST message
specification\footnote{\url{http://www.cmegroup.com/confluence/download/attachments/62816403/Legacy/\%2520FIX-FAST/\%2520Market/\%2520Dta/\%2520Message/\%2520Specification.pdf?version=2\&modificationDate=1452872661000\&api=v2}}.
We have made the translation scripts used in this analysis freely
available\footnote{\url{https://github.com/Matt-Brigida/CME-FIX-FAST-Translator}}.

In the following charts and analysis it is helpful to note the
difference between \texttt{clock} and \texttt{market\ time}. When
considering the nanosecond (one-billionth of a second) level, the market
has long periods of inactivity interspersed with periods of activity.
Our data set only contains these periods of activity (and of course the
length of time since the previous period of activity). Otherwise we
would require a time series of 1 billion data points to analyze each
second.

\hypertarget{methods}{%
\section{Methods}\label{methods}}

Our challenge is that of \emph{unsurpervised learning}---we are
attempting to identify a state without training data providing the
states for a sample of data. A classic problem of this type in the
economics literature is to determine if the economy is in an expansion
or recession. In this expansion/recession analysis Markov
regime-switching regressions are used (see for example the method
employed by the US Federal Reserve). We'll use a similar approach in our
analysis to determine periods of stable, and unstable, liquidity driven
by algorithmic activity. Our exact model is outlined below.

We measure liquidity on each side of the book as the amount of ES that
can be bought within one point of the present bid-offer midpoint. One
point is equivalent to 4 ticks (so maximum the inside quote and 3
additional levels of the book). Results below are for the November 9,
2016 trading day, which is the most likely to exhibit algorithmic
trading activity due to the large public release of information, and the
consequent portfolio rebalancing and increased trade volume.

\hypertarget{markov-switching-model}{%
\subsection{Markov-Switching Model}\label{markov-switching-model}}

There is no test for the proper number of states in a multiple state
model. We thus estimate an increasing number of states and let the
interpretation of the results and standard tests of the residuals, in
each state, to guide us to finding a state consistent with algorithmic
activity.

The two-state version of our model is:

\[
Liq_t =
\begin{cases}
\alpha_1 + \beta_{11} Liq_{t-1} + \beta_{12} \Delta BAM + \epsilon_1, \ \ \ \epsilon_1 \sim N(0, \sigma_1)    \\
\alpha_2 + \beta_{21} Liq_{t-1} + \beta_{22} \Delta BAM + \epsilon_2, \ \ \ \epsilon_2 \sim N(0, \sigma_2)    \\
\end{cases}
\]

\[
P(s_t = j | s_{t-1} = i) = p_{ij} \\ \text{for} \ \ i,j \in {1,2} \ and \ \sum_{j=1}^2p_{ij} = 1
\]

where \(Liq_{t-1}\) is the liquidity in the previous period and
\(\Delta BAM\) is the most recent change in the bid-ask midpoint. There
are two states, denoted by \(s_1\) and \(s_2\), and \(p_{ij}\) denotes
the probability that the state is \(j\) given the state was \(i\) in the
previous period. We estimate the model via the Hamilton Filter with a
custom implementation in C++ due to the large number of points in our
time series.

Similar to the bid and ask volatility estimate in Hasbrouck (2018), we
estimate the model for the bid and ask sides of the book separately.
This is because the rapid deviations from a mean liquidity value, which
we are attempting to identify, largely affect one side of the book, and
so are more likely to be an artifact of the trading process rather than
due to fundamental information. Nonetheless, modeling the entire book
(bid and ask sides jointly) would include more information in the
parameter estimates, such as spillover effects. However this would
increase the time required to estimate parameters as well as the time it
takes to create a state prediction. Since the algorithm must be very
quick to be useful, we err on the side of speed relative to the benefit
of the information in both sides of the spread.

\hypertarget{two-states}{%
\subsubsection{Two States}\label{two-states}}

The two-state model is picking up states of changing liquidity and
stable liquidity. In both the bid and offer models, the first state had
a coefficient of 1 on the previous liquidity, and a small residual
standard deviation. This state is consistent with no public or private
information being incorporated into prices, and little in the market
changing.

The second state, which has a higher residual variance, exhibits
evidence of changing liquidity. However the coefficient on previous
liquidity, and the intercept are significantly different between the two
models. Accordingly, state 2 may be driven by liquidity changing for
various reasons. These results motivate a 3 state model where we
differentiate the state with changing liquidity into two states---one
representing changing liquidity due to HFT activity:

\begin{itemize}
\tightlist
\item
  Stable liquidity
\item
  Normal changing liquidity
\item
  Changing liquidity due to HFT
\end{itemize}

\hypertarget{bid}{%
\paragraph{Bid}\label{bid}}

\[
Liq_t =
\begin{cases}
0.00 + 1.00 Liq_{t-1} + 0.09 \Delta BAM + \epsilon_2, \ \ \ \epsilon_2 \sim N(0, 0.002)    \\
-0.83 + 0.49 Liq_{t-1} - 0.06 \Delta BAM + \epsilon_1, \ \ \ \epsilon_1 \sim N(0, 0.47)    \\
\end{cases}
 \]

\begin{figure}
\centering
\includegraphics{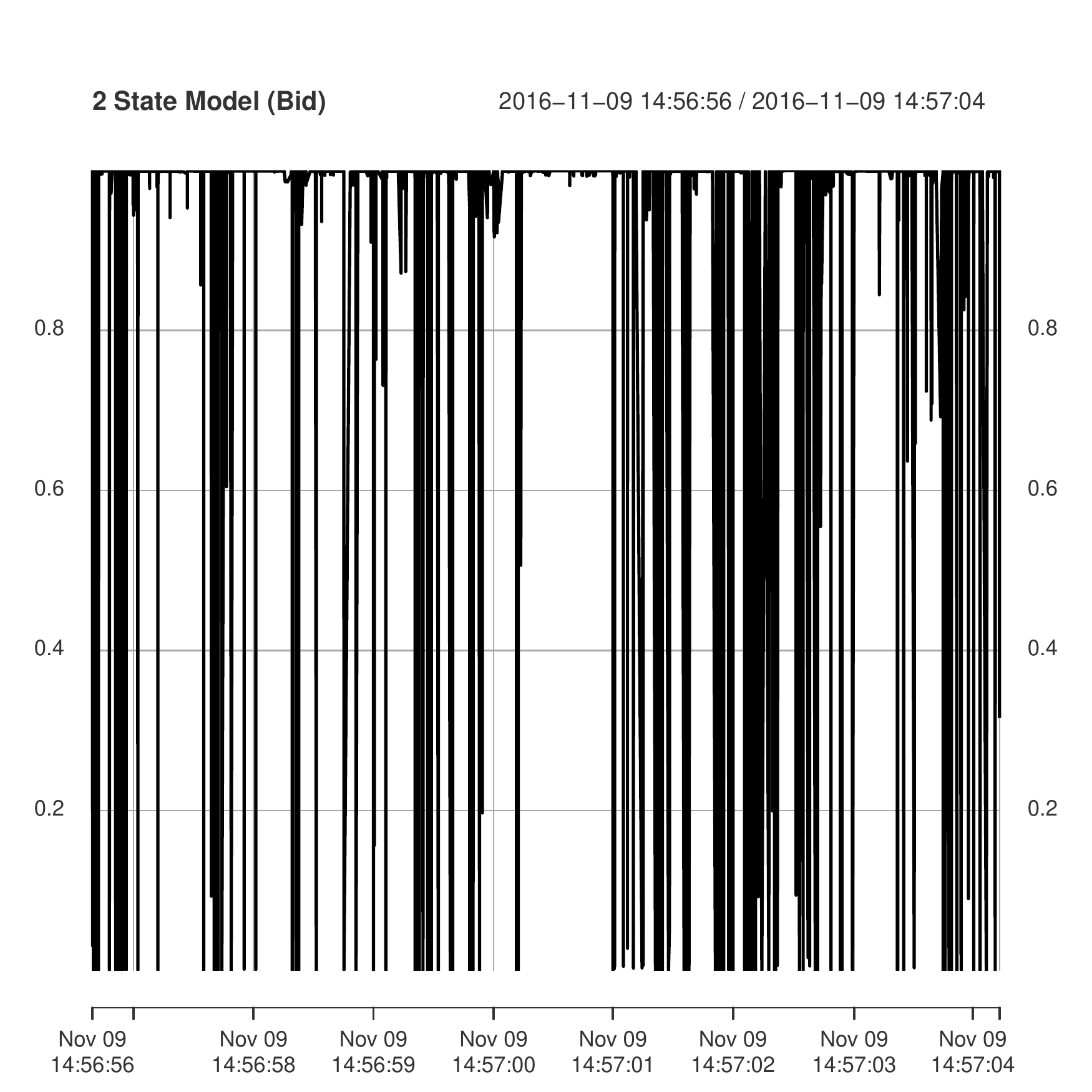}
\caption{Two state Markov-Switching model of liquidity available at the
bid.}
\end{figure}

\hypertarget{offer}{%
\paragraph{Offer}\label{offer}}

\[
Liq_t =
\begin{cases}
0.42 + 1.33 Liq_{t-1} - 0.12 \Delta BAM + \epsilon_1, \ \ \ \epsilon_1 \sim N(0, 0.42)    \\
0.00 + 1.00 Liq_{t-1} + 0.16 \Delta BAM + \epsilon_2, \ \ \ \epsilon_2 \sim N(0, 0.001)    \\
\end{cases}
 \]

\begin{figure}
\centering
\includegraphics{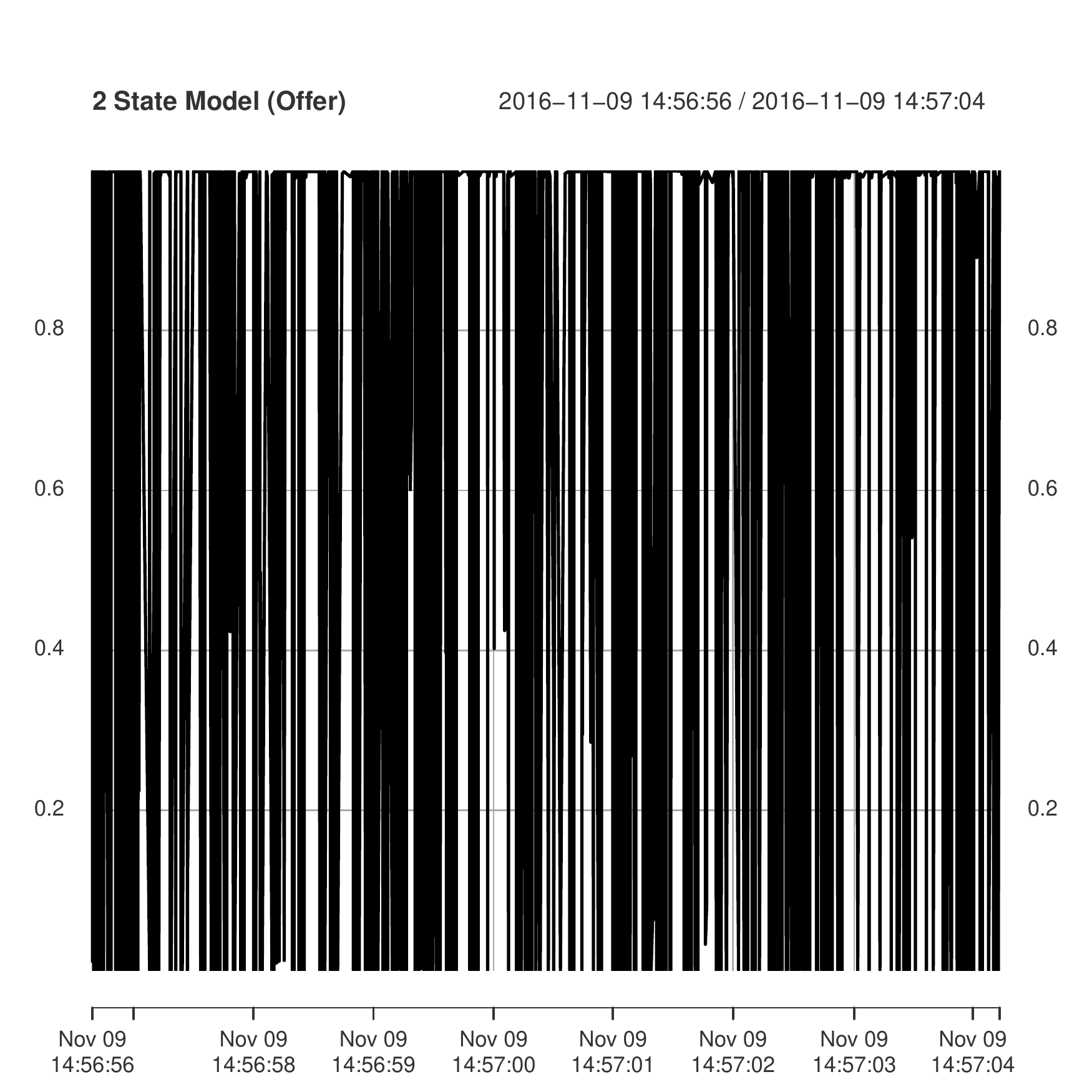}
\caption{Two state Markov-Switching model of liquidity available at the
offer.}
\end{figure}

\hypertarget{three-states}{%
\subsubsection{Three States}\label{three-states}}

The first state in the 3 state model again exhibits stable liquidity.
The following two states exhibit varying volatility which is driven by
different factors. In state 2 liquidity is driven by a change in the
bid-ask midpoint. This is consistent with liquidity provision in
reaction to a movement in the market---posssibly driven by new
information.

In state 3, however, a change in the bid-ask midpoint has no effect on
liquidity. Similarly previous liquidity explains only a quarter to a
third of present liquidity, and the variance of the residual is the
highest in state 3. If there is HFT activity present, it is most likely
within state 3. Note, these results are consistent across both bids and
asks. Lastly, we'll estimate the parameters of a four state model to see
if state 3 is a composite of other states.

\hypertarget{bid-1}{%
\paragraph{Bid}\label{bid-1}}

\[
Liq_t =
\begin{cases}
-0.00 + 1 Liq_{t-1} -0.12 \Delta BAM + \epsilon_1 \\
-0.09 + 0.22 Liq_{t-1} + 1.02 \Delta BAM + \epsilon_2 \\
-0.01 + 0.32 Liq_{t-1} + 0.004 \Delta BAM + \epsilon_3 \\
\end{cases}
\]

\[
\texttt{where}
\begin{cases}
\epsilon_1 \sim N(0, 0.004)    \\
\epsilon_2 \sim N(0, 0.2929)    \\
\epsilon_3 \sim N(0, 0.40)    \\
\end{cases}
\]

\begin{figure}
\centering
\includegraphics{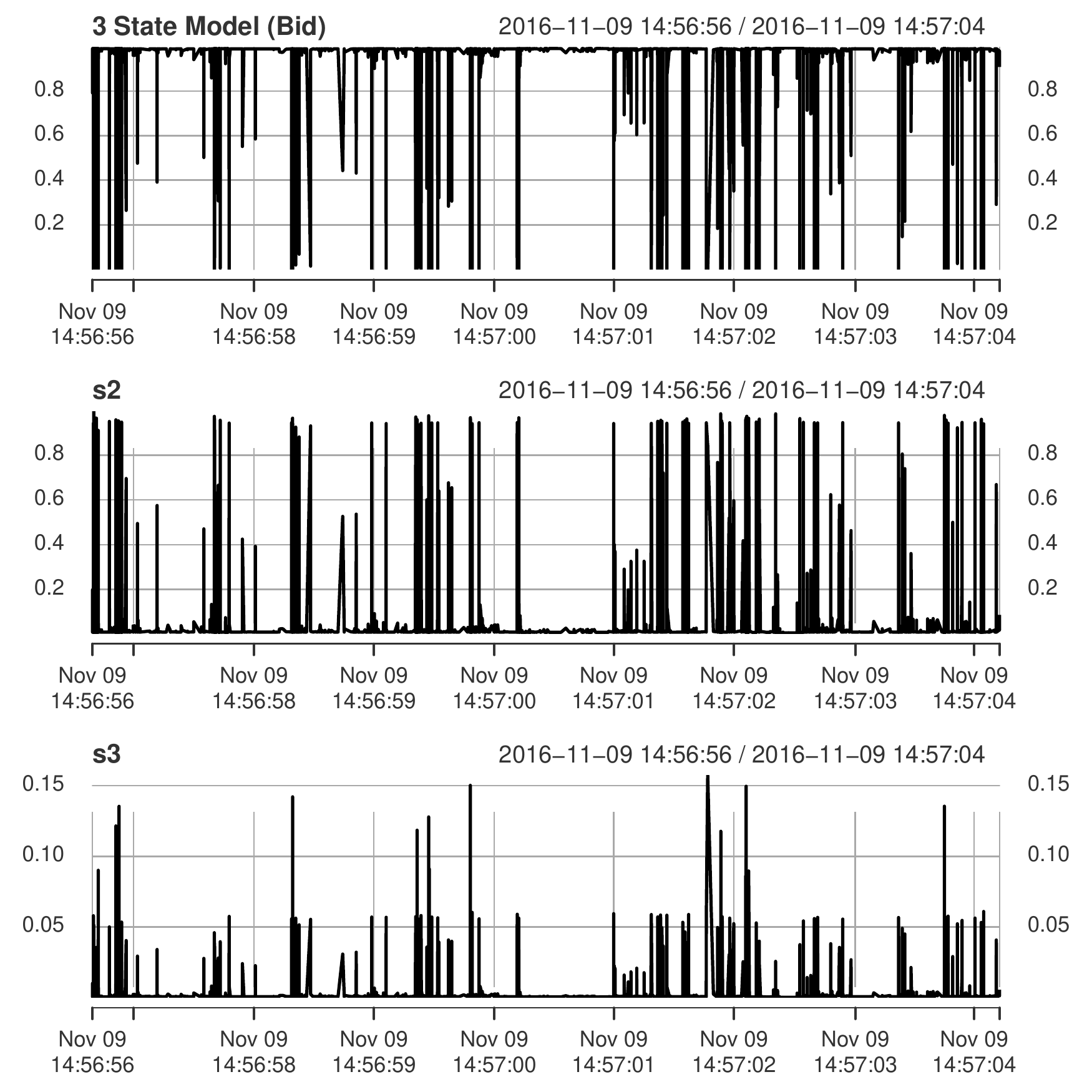}
\caption{Three state Markov-Switching model of liquidity available at
the bid.}
\end{figure}

\hypertarget{offer-1}{%
\paragraph{Offer}\label{offer-1}}

\[
Liq_t =
\begin{cases}
-0.00 +  1 Liq_{t-1} - 0.10 \Delta BAM + \epsilon_1 \\
0.38 +  -0.03 Liq_{t-1} +  0.81 \Delta BAM + \epsilon_2 \\
0.12 +  0.25 Liq_{t-1} +  0.01 \Delta BAM + \epsilon_3 \\
\end{cases}
\]

\[
\texttt{where}
\begin{cases}
\epsilon_1 \sim N(0, 0.004)    \\
\epsilon_2 \sim N(0, 0.078)    \\
\epsilon_3 \sim N(0, 0.90)    \\
\end{cases}
\]

\begin{figure}
\centering
\includegraphics{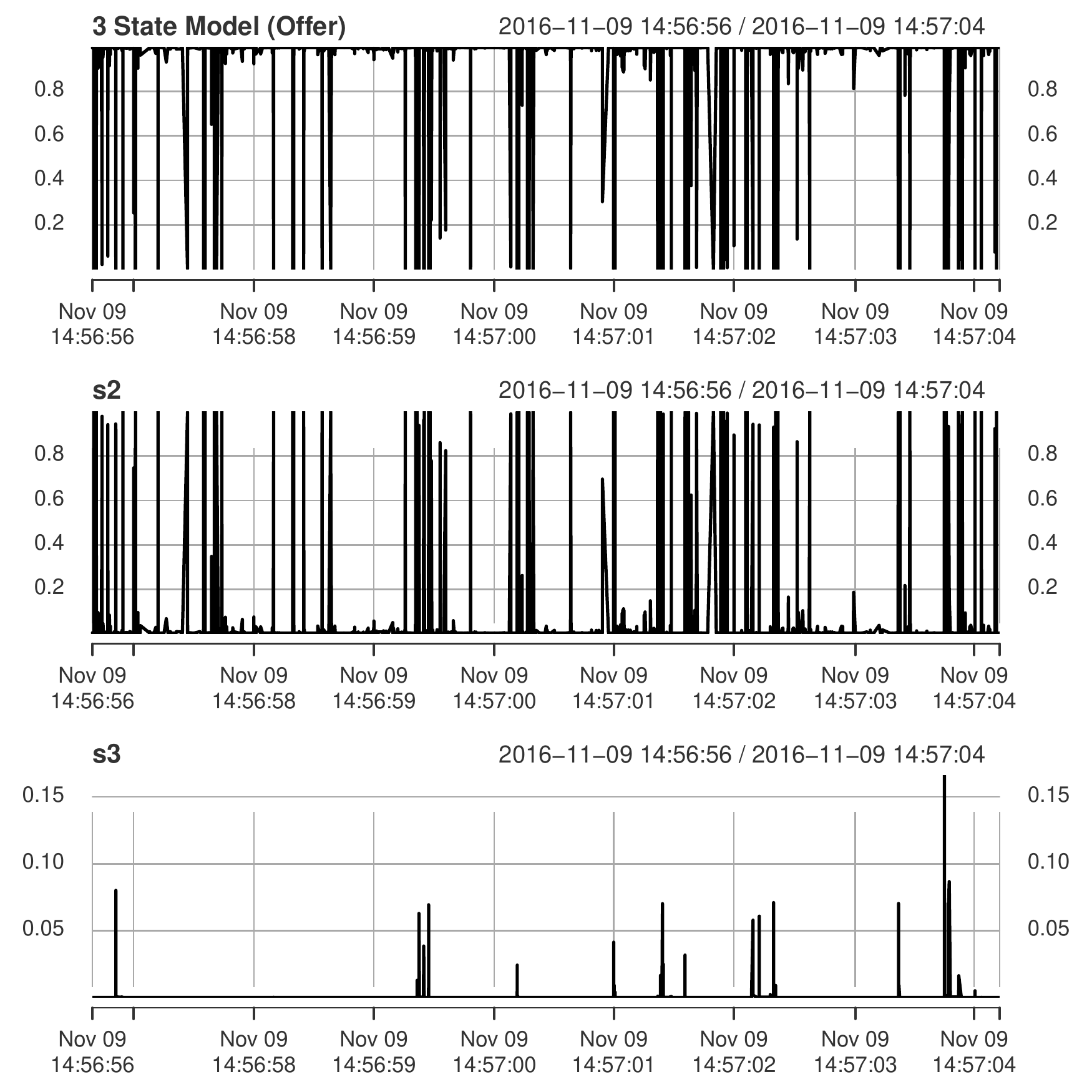}
\caption{Three state Markov-Switching model of liquidity available at
the offer.}
\end{figure}

\hypertarget{four-states}{%
\subsubsection{Four States}\label{four-states}}

Similar to the three-state equation, the first two states represent
stable liquidity, and changing liquidity driven by changes in the
bid-ask midpoint. State 3 exhibits negative relationships between
previous and present liquidity. The standard deviation of the error term
is moderately high in this state, however it is about a quarter to a
third of the standard deviation of the error term in state 4.

State 4 is most consistent with the type of HFT activity we are trying
to identify. In state 4 liquidity remains constant with substantial
variability around the stable mean liquidity amount.

\hypertarget{bid-2}{%
\paragraph{Bid}\label{bid-2}}

\[
Liq_t =
\begin{cases}
0.0024 + 0.9983 Liq_{t-1} + 0.1319 \Delta BAM + \epsilon_1 \\
-0.0594 - 0.3211 Liq_{t-1} + 0.8524 \Delta BAM + \epsilon_2 \\
0.3796 - 0.0636 Liq_{t-1} - 0.1802 \Delta BAM + \epsilon_3 \\
-0.1626 + 0.9469 Liq_{t-1} + 0.0791 \Delta BAM + \epsilon_4 \\
\end{cases}
\]

\[
\texttt{where}
\begin{cases}
\epsilon_1 \sim N(0, 0.0077)    \\
\epsilon_2 \sim N(0, 0.2901)    \\
\epsilon_3 \sim N(0, 0.2409)    \\
\epsilon_4 \sim N(0, 0.6580)    \\
\end{cases}
\]

\begin{figure}
\centering
\includegraphics{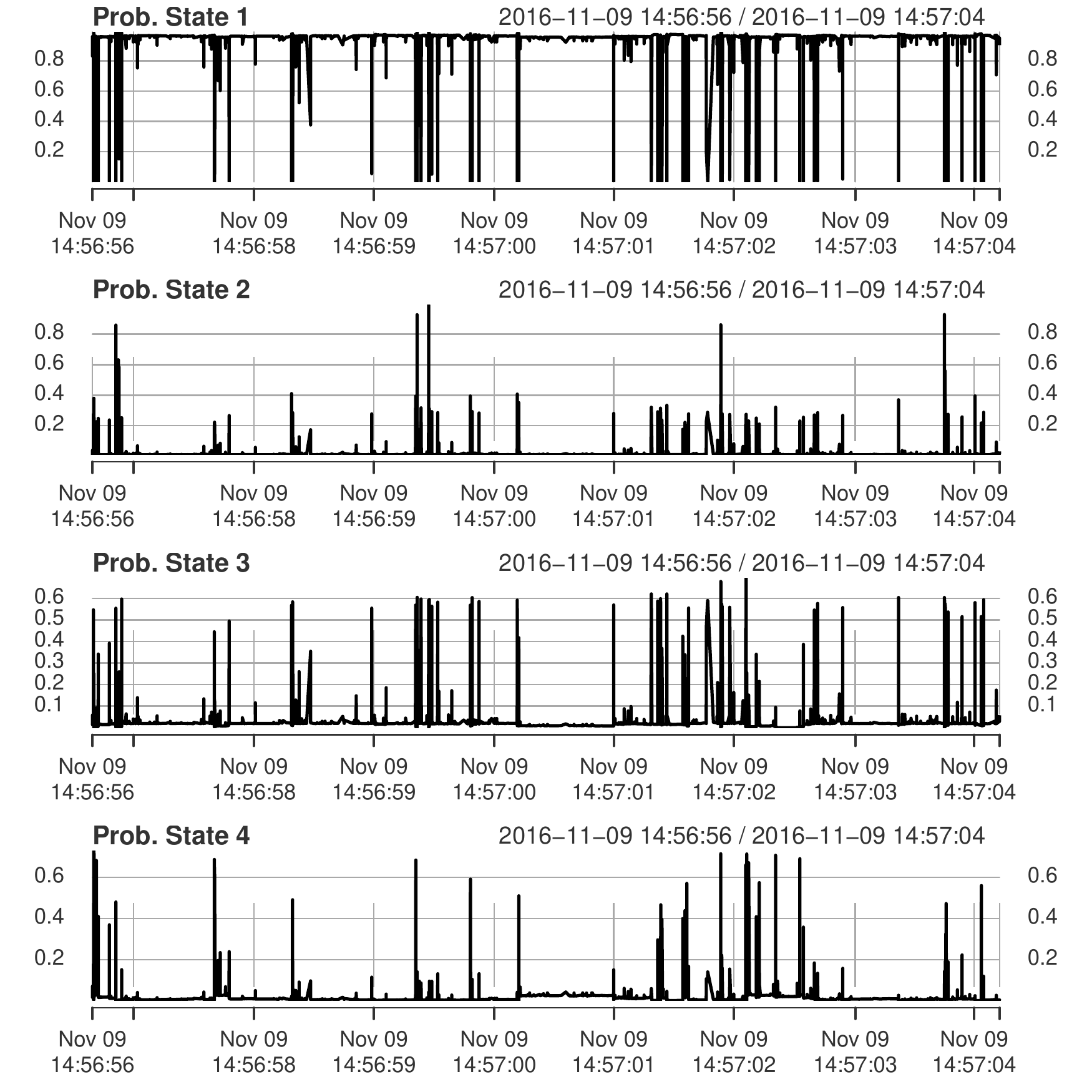}
\caption{Four state Markov-Switching model of liquidity available at the
bid.}
\end{figure}

\hypertarget{offer-2}{%
\paragraph{Offer}\label{offer-2}}

\[
Liq_t =
\begin{cases}
0.0000 +  1.0000 Liq_{t-1} - 0.2681  \Delta BAM + \epsilon_1 \\
-0.0055 +  0.9949 Liq_{t-1} - 1.1200 \Delta BAM + \epsilon_2 \\
-1.1325 -  0.3480 Liq_{t-1} - 0.0122  \Delta BAM + \epsilon_3 \\
-0.0048 + 1.0051 Liq_{t-1} -  0.5034 \Delta BAM + \epsilon_4 \\
\end{cases}
\]

\[
\texttt{where}
\begin{cases}
\epsilon_1 \sim N(0, 0.0000)    \\
\epsilon_2 \sim N(0, 0.0153)    \\
\epsilon_3 \sim N(0, 0.1400)    \\
\epsilon_4 \sim N(0, 0.6207)    \\
\end{cases}
\]

\begin{figure}
\centering
\includegraphics{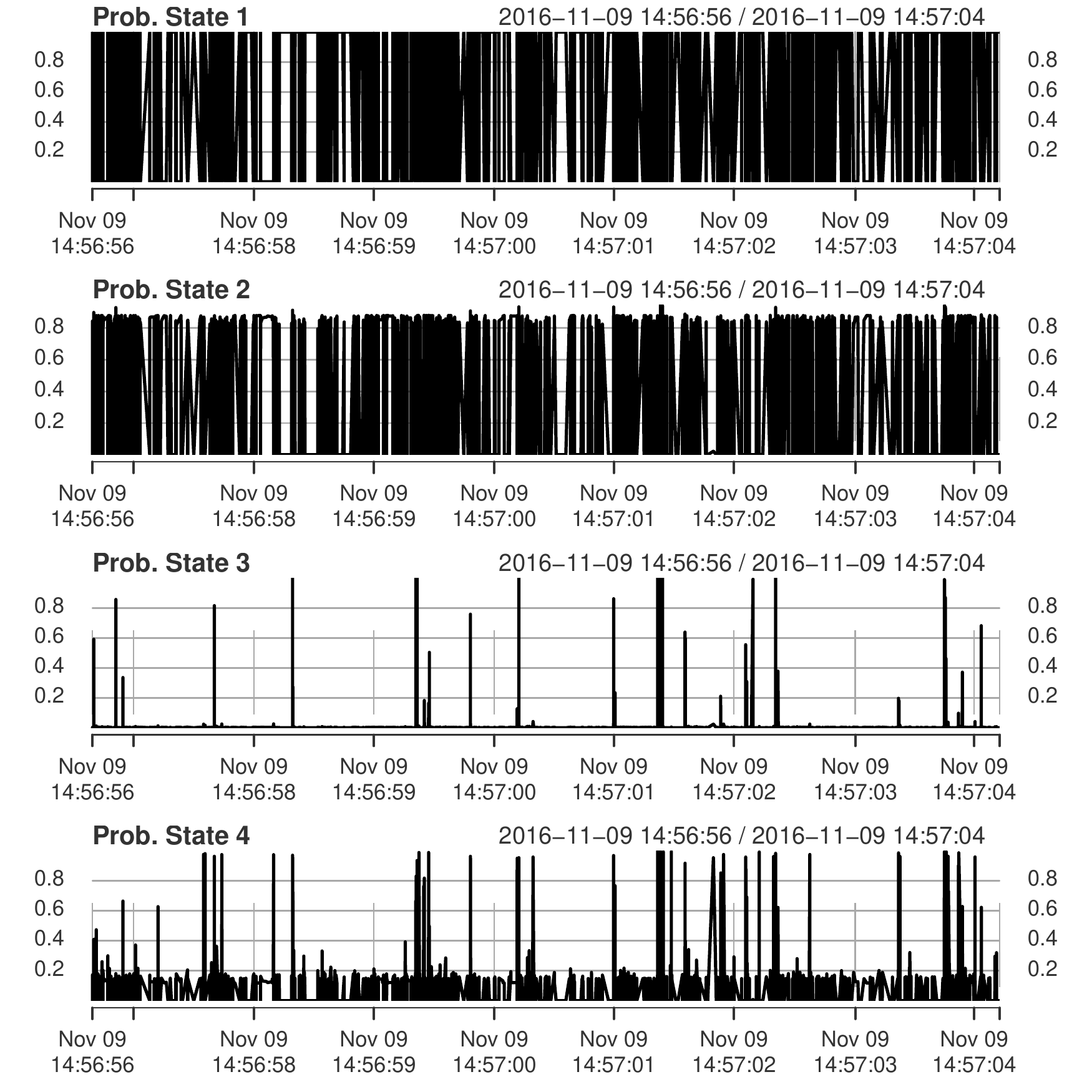}
\caption{Four state Markov-Switching model of liquidity available at the
offer.}
\end{figure}

Given the above estimates, and using 0.2 as our signal threshold for
state 4, the signal fires, on average, 0.636 times per second on the bid
side of the orderbook. On the ask side of the orderbook the signal fires
10.59 times per second on average. Assuming a 10 millisecond delay each
time the signal fires, this implies the signal duration of 0.636\% and
10.59\% of the trading day on the bid and ask side of the orderbook
respectively. This duration range is reasonable given anecdotal accounts
of the pervasiveness of high-frequency trading in markets, such as
Hendershott, Jones, and Menkveld (2011) which reported that as much as
73\% of volume in US markets was due to high-frequency trading.

In tables 3 and 4 in the appendix we provide parameter estimates for the
4 state model, along with signal duration estimates, for the entire week
(November 7 through 11, 2016). The parameter estimates are very similar
across days for each side of the orderbook. Further the signal durations
are also similar with the exception of the offer side of the book on the
November 9th trading day. The large release of public information
occurred on November 9th, and this orderbook asymmetry with regards to
algorithmic activity is consistent with Hasbrouck (2018).

\begin{longtable}[]{@{}llll@{}}
\caption{\textbf{Bid side of the orderbook}. Below are coefficient
estimates from the Markov-switching regressions. The standard errors are
next to the coefficient in parentheses. The coefficients were estimated
using the nanosecond time-stamped orderbook ranging from 6:00 PM EST on
November 8, 2016 to 5:00 PM EST on November 9, 2016. There are 9,965,673
changes to the orderbook for this period.}\tabularnewline
\toprule
Coefficient & Two-State & Three-State & Four-State\tabularnewline
\midrule
\endfirsthead
\toprule
Coefficient & Two-State & Three-State & Four-State\tabularnewline
\midrule
\endhead
\(\alpha_1\) & 0.00(0.0000) & -0.00(0.0000) &
0.00(0.0000)\tabularnewline
\(\alpha_2\) & -0.83(0.0007) & -0.09(0.0250) &
-0.05(0.0033)\tabularnewline
\(\alpha_3\) & & -0.01(0.0140) & 0.37(0.0025)\tabularnewline
\(\alpha_4\) & & & -0.16(0.0018)\tabularnewline
\(\beta_{11}\) & 1.00(0.0000) & 1.00(0.0000) &
0.99(0.0000)\tabularnewline
\(\beta_{12}\) & 0.09(0.1369) & -0.12(0.259) &
0.13(0.4226)\tabularnewline
\(\beta_{21}\) & 0.49(0.0004) & 0.22(0.003) &
-0.32(0.0110)\tabularnewline
\(\beta_{22}\) & -0.06(0.0075) & 1.02(0.670) &
0.85(1.1350)\tabularnewline
\(\beta_{31}\) & & 0.32(0.000) & -0.06(0.0050)\tabularnewline
\(\beta_{32}\) & & 0.00(0.0000) & -0.18(0.8833)\tabularnewline
\(\beta_{41}\) & & & 0.94(0.0001)\tabularnewline
\(\beta_{42}\) & & & 0.07(0.0933)\tabularnewline
\(\sigma_1\) & 0.00(0.0490) & 0.00(0.0661) & 0.00(0.0000)\tabularnewline
\(\sigma_2\) & 0.47(0.0002) & 0.29(0.0024) & 0.29(0.0044)\tabularnewline
\(\sigma_3\) & & 0.40(0.0001) & 0.24(0.0033)\tabularnewline
\(\sigma_4\) & & & 0.65(0.0008)\tabularnewline
\bottomrule
\end{longtable}

\begin{longtable}[]{@{}llll@{}}
\caption{\textbf{Ask side of the orderbook}. Below are coefficient
estimates from the Markov-switching regressions. The standard errors are
next to the coefficient in parentheses. The coefficients were estimated
using the nanosecond time-stamped orderbook ranging from 6:00 PM EST on
November 8, 2016 to 5:00 PM EST on November 9, 2016. There are 9,965,673
changes to the orderbook for this period.}\tabularnewline
\toprule
Coefficient & Two-State & Three-State & Four-State\tabularnewline
\midrule
\endfirsthead
\toprule
Coefficient & Two-State & Three-State & Four-State\tabularnewline
\midrule
\endhead
\(\alpha_1\) & 0.42(0.0000) & -0.00(0.0000) &
0.00(0.0000)\tabularnewline
\(\alpha_2\) & 0.00(0.0041) & 0.38(0.0141) &
-0.00(0.0970)\tabularnewline
\(\alpha_3\) & & 0.12(0.0196) & -1.13(0.7924)\tabularnewline
\(\alpha_4\) & & & -0.00(0.02269)\tabularnewline
\(\beta_{11}\) & 1.33(0.4078) & 1.00(0.0000) &
1.00(0.0083)\tabularnewline
\(\beta_{12}\) & -0.12(0.0059) & -0.10(0.2259) &
-0.26(0.1421)\tabularnewline
\(\beta_{21}\) & 1.00(0.0000) & -0.03(0.0192) &
0.99(0.0001)\tabularnewline
\(\beta_{22}\) & 0.16(0.0009) & 0.81(1.5312) &
1.12(0.0018)\tabularnewline
\(\beta_{31}\) & & 0.25(0.0027) & -0.34(0.0990)\tabularnewline
\(\beta_{32}\) & & 0.01(1.3956) & -0.01(0.6147)\tabularnewline
\(\beta_{41}\) & & & 1.00(0.2876)\tabularnewline
\(\beta_{42}\) & & & -0.50(0.9778)\tabularnewline
\(\sigma_1\) & 0.42(0.0011) & 0.00(0.0000) & 0.00(0.0000)\tabularnewline
\(\sigma_2\) & 0.00(0.0327) & 0.07(0.0101) & 0.01(0.0626)\tabularnewline
\(\sigma_3\) & & 0.90(0.0002) & 0.14(0.0115)\tabularnewline
\(\sigma_4\) & & & 0.62(0.0004)\tabularnewline
\bottomrule
\end{longtable}

\hypertarget{conclusion}{%
\section{Conclusion}\label{conclusion}}

In this analysis we have used Markov-Switching regression models to
identify the presence of high-frequency traders who are rapidly changing
volatility. Using a model with four states, we identify a state with a
stable mean liquidity, but substantial variability in liquidity around
the mean. That is there is rapidly changing liquidity, which does not
affect overall liquidity or the price.

Since trading in this state benefits high-frequency traders at the
expense of slower retail order flow, a transition to this state can
serve as a signal to delay slower traders' orders. The delay being mere
tens of milliseconds, it will not be perceptible to the typical trader.
And while this may save each trade a small amount, in aggregate such a
delayed order type would provide substatial savings across all
non-high-frequency traders. Delaying orders due to the signal can be
offered to retail traders through a particular order type. A similar
strategy is used by the IEX's `crumbling quote' order.

\clearpage

\hypertarget{appendix}{%
\section{Appendix}\label{appendix}}

In tables 3 and 4 below are parameter estimates from the following
4-state Markov-Switching model.

\begin{table}[htbp]
\caption{Parameter estimates from a 4-state Markov-switching model on the liquidity available on the bid side of the orderbook.  There are 2,917,466 entries to the book over the Nov. 7 trading day.  There are 3,502,097 book entries on Nov. 8.  There are 9,965,673 book entries on Nov. 9, which is the trading day over which the results of the election were announced.  There were 7,346,604 book entries on Nov. 10, and 4,905,882 on Nov. 11.  The duration of the signal (Sig. Dur.) was calculated assuming a 10 millisecond delay for each signal, and a 0.2 threshold for the signal generation.}
\centering
\begin{tabular}{lrrrrr}
\hline
Coefficient & Nov. 7 & Nov. 8 & Nov. 9 & Nov. 10 & Nov. 11\\
\hline
$\alpha_1$ & 0.0065 & 0.0431 & 0.0024 & -0.0656 & -0.0010\\
$\alpha_2$ & -0.1132 & -0.1694 & -0.0594 & -0.4849 & -0.4917\\
$\alpha_3$ & 0.1121 & 0.3509 & 0.3796 & 0.3917 & 0.2975\\
$\alpha_4$ & -0.2210 & -0.1783 & -0.1626 & -0.2966 & -0.3563\\
$\beta_{11}$ & 1.0004 & 0.8102 & 0.9983 & 0.9500 & 1.0057\\
$\beta_{12}$ & -0.1579 & 0.0754 & 0.1319 & 0.0174 & 0.2716\\
$\beta_{21}$ & 0.1741 & 0.0168 & -0.3211 & 0.2004 & 0.2565\\
$\beta_{22}$ & 0.9270 & 0.8738 & 0.8524 & 0.8375 & 0.4647\\
$\beta_{31}$ & 0.0628 & 0.1031 & -0.0636 & 0.0132 & 0.0101\\
$\beta_{32}$ & -0.1324 & -0.1707 & -0.1802 & -0.1676 & -0.3864\\
$\beta_{41}$ & 0.6621 & 0.6239 & 0.9469 & 0.4445 & 1.1467\\
$\beta_{42}$ & 0.1151 & 0.0910 & 0.0791 & -0.0752 & -0.2919\\
$\sigma_1$ & 0.0221 & 0.0912 & 0.0077 & 0.0109 & 0.0219\\
$\sigma_2$ & 0.0920 & 0.1716 & 0.2901 & 0.4268 & 0.6963\\
$\sigma_3$ & 0.1701 & 0.0787 & 0.2409 & 0.0769 & 0.1083\\
$\sigma_4$ & 0.6386 & 0.6873 & 0.6580 & 0.4252 & 0.3709\\
\hline
Log Lik. & 4880164 & 117503.2 & 16693395 & 20395.45 & 249944.1\\
Sig. Dur. & 0.736\% & 0.020\% & 0.636\% & 0.000\% & 0.000\%\\
\hline
\end{tabular}
\end{table}

\begin{table}[htbp]
\caption{Parameter estimates from a 4-state Markov-switching model on the liquidity available on the offer side of the orderbook.  There are 2,917,466 entries to the book over the Nov. 7 trading day.  There are 3,502,097 book entries on Nov. 8.  There are 9,965,673 book entries on Nov. 9, which is the trading day over which the results of the election were announced.  There were 7,346,604 book entries on Nov. 10, and 4,905,882 on Nov. 11. The duration of the signal (Sig. Dur.) was calculated assuming a 10 millisecond delay for each signal, and a 0.2 threshold for the signal generation.}
\centering
\begin{tabular}{lrrrrr}
\hline
Coefficient & Nov. 7 & Nov. 8 & Nov. 9 & Nov. 10 & Nov. 11\\
\hline
$\alpha_1$ & 0.0000 & 0.0000 & 0.0000 & 0.0000 & 0.0000\\
$\alpha_2$ & -0.0007 & 0.0008 & -0.0055 & 0.0011 & -0.0051\\
$\alpha_3$ & -1.1325 & -1.1374 & -1.1325 & -0.1314 & -1.1329\\
$\alpha_4$ & -0.0042 & 0.0052 & -0.0048 & -0.0060 & 0.0015\\
$\beta_{11}$ & 1.0054 & 1.0051 & 1.0049 & 1.0059 & 0.9979\\
$\beta_{12}$ & -0.2681 & -0.2707 & -0.2681 & -0.2643 & -0.0033\\
$\beta_{21}$ & 0.9960 & 0.9977 & 0.9949 & 0.9991 & 0.9955\\
$\beta_{22}$ & -1.1161 & -1.1034 & -1.1200 & -1.1195 & -1.291\\
$\beta_{31}$ & 0.3465 & 0.3481 & 0.3480 & 0.3487 & 0.3414\\
$\beta_{32}$ & -0.0121 & -0.0153 & -0.0122 & -0.0082 & -0.0064\\
$\beta_{41}$ & 1.2411 & 1.2369 & 1.0086 & 1.2357 & 1.0148\\
$\beta_{42}$ & -0.5031 & -0.5057 & -0.5034 & -0.5006 & -0.5136\\
$\sigma_1$ & 0.0000 & 0.0000 & 0.0000 & 0.0000 & 0.0000\\
$\sigma_2$ & 0.0016 & 0.0018 & 0.0153 & 0.0030 & 0.0039\\
$\sigma_3$ & 0.1400 & 0.1400 & 0.1400 & 0.1400 & 0.1400\\
$\sigma_4$ & 0.6217 & 0.6369 & 0.6207 & 0.6316 & 0.6165\\
\hline
Log Lik. & 841089.4 & 5872364 & 45918365 & 5637773 & 839429.5\\
Sig. Dur. & 0.403\% & 0.797\% & 10.59\% & 0.704\% & 1.206\%\\
\hline
\end{tabular}
\end{table}

\clearpage

\hypertarget{references}{%
\section*{References}\label{references}}
\addcontentsline{toc}{section}{References}

\hypertarget{refs}{}
\begin{CSLReferences}{1}{0}
\leavevmode\hypertarget{ref-bishop2017evolution}{}%
Bishop, Allison. 2017. ``The Evolution of the Crumbling Quote Signal.''
\emph{Available at SSRN 2956535}.

\leavevmode\hypertarget{ref-COPELAND_1983}{}%
Copeland, Thomas E., and Dan Galai. 1983. ``Information Effects on the
Bid-Ask Spread.'' \emph{The Journal of Finance} 38 (5): 1457--69.
\url{https://doi.org/10.1111/j.1540-6261.1983.tb03834.x}.

\leavevmode\hypertarget{ref-EASLEY_1992}{}%
Easley, Maureen, David And O'Hara. 1992. ``Time and the Process of
Security Price Adjustment.'' \emph{The Journal of Finance} 47 (2):
577--605. \url{https://doi.org/10.1111/j.1540-6261.1992.tb04402.x}.

\leavevmode\hypertarget{ref-French_1986}{}%
French, Kenneth R., and Richard Roll. 1986. ``Stock Return Variances.''
\emph{Journal of Financial Economics} 17 (1): 5--26.
\url{https://doi.org/10.1016/0304-405x(86)90004-8}.

\leavevmode\hypertarget{ref-Glosten_1985}{}%
Glosten, Lawrence R., and Paul R. Milgrom. 1985. ``Bid, Ask and
Transaction Prices in a Specialist Market with Heterogeneously Informed
Traders.'' \emph{Journal of Financial Economics} 14 (1): 71--100.
\url{https://doi.org/10.1016/0304-405x(85)90044-3}.

\leavevmode\hypertarget{ref-hasbrouck2003intraday}{}%
Hasbrouck, Joel. 2003. ``Intraday Price Formation in Us Equity Index
Markets.'' \emph{The Journal of Finance} 58 (6): 2375--2400.

\leavevmode\hypertarget{ref-hasbrouck2018high}{}%
---------. 2018. ``High-Frequency Quoting: Short-Term Volatility in Bids
and Offers.'' \emph{Journal of Financial and Quantitative Analysis} 53
(2): 613--41.

\leavevmode\hypertarget{ref-hendershott2011does}{}%
Hendershott, Terrence, Charles M Jones, and Albert J Menkveld. 2011.
``Does Algorithmic Trading Improve Liquidity?'' \emph{The Journal of
Finance} 66 (1): 1--33.

\leavevmode\hypertarget{ref-Kyle_1985}{}%
Kyle, Albert S. 1985. ``Continuous Auctions and Insider Trading.''
\emph{Econometrica} 53 (6): 1315. \url{https://doi.org/10.2307/1913210}.

\leavevmode\hypertarget{ref-Lee_1993}{}%
Lee, Charles M. C., Belinda Mucklow, and Mark J. Ready. 1993. ``Spreads,
Depths, and the Impact of Earnings Information: An Intraday Analysis.''
\emph{Review of Financial Studies} 6 (2): 345--74.
\url{https://doi.org/10.1093/rfs/6.2.345}.

\end{CSLReferences}

\end{document}